**Effects of pressure on the magnetostructural and magnetocaloric properties of isostructurally alloyed $(MnNiSi)_{1-x}(FeCoGe)_x$**

Tapas Samanta[1*], Daniel L. Lepkowski[1], Ahmad Us Saleheen[1], Alok Shankar[1], Joseph Prestigiacomo[1], Igor Dubenko[2], Abdiel Quetz[2], Iain W. H. Oswald[3], Gregory T. McCandless[3], Julia Y. Chan[3], Philip W. Adams[1], David P. Young[1], Naushad Ali[2], Shane Stadler[1]

[1]Department of Physics & Astronomy, Louisiana State University, Baton Rouge, LA 70803 USA
[2]Department of Physics, Southern Illinois University, Carbondale, IL 62901 USA
[3]Department of Chemistry, The University of Texas at Dallas, Richardson, TX 75080 USA
[*]Correspondence to: tsamanta@lsu.edu

The isostructural alloying of two compounds with extremely different magnetic and thermo-structural properties has resulted in a new system, $(MnNiSi)_{1-x}(FeCoGe)_x$, that exhibits extraordinary magnetocaloric properties with an acute sensitivity to applied hydrostatic pressure ($P$). Application of hydrostatic pressure shifts the first-order phase transition to lower temperature ($\Delta T$=–41 K with $P$=3.43 kbar) but preserves the giant value of isothermal entropy change ($-\Delta S^{max}$=143.7 J/kg K for a field change of $\Delta B$=5 T at atmospheric pressure). Together with the magnetic field, this pressure-induced temperature shift can be used to significantly increase the effective relative cooling power.

In recent years, considerable attention has been devoted to studies of Mn-based MnTX (T = Co, Ni and X = Ge, Si) systems due to their temperature-induced magnetostructural transitions (MST) that result in shape memory phenomena, giant magnetocaloric effects (MCE), and volume anomalies near room temperature [1-9]. Some also behave as strongly-correlated electron systems in the proximity of a noncollinear ferromagnetic state [10]. In particular, the coincidence of magnetic and structural transitions near room temperature induced by properly tuning the stoichiometry and chemical composition along with the associated large MCE, make these systems of great interest in the field of magnetocalorics

A strong coupling of magnetic and structural degrees of freedom often results in a giant MCE, as observed in many well-known magnetocaloric materials in the vicinity of a magnetostructural transition (MST), accompanied by changes in crystal symmetry or volume. A large structural entropy change associated with a significant volume change due to the structural transition can enhance the total entropy change in MnTX systems in comparison to the other well-known giant magnetocaloric materials. Pressure is a controllable external parameter that can affect the structural entropy change of a system and, as a result, a pressure-induced enhancement of magnetocaloric properties could be expected in some MnTX systems. Recent reports on hydrostatic-pressure studies also indicate the possibility of applying pressure to improve the magnetocaloric properties by demonstrating a large isothermal entropy change [11, 12].

Here, we report the discovery of a system, $(MnNiSi)_{1-x}(FeCoGe)_x$, in which applied hydrostatic pressure shifts the temperature of the phase transition responsible for the MCE, providing a method to tune it over a broad temperature range. Importantly, the applied pressure not only shifts the transition temperatures but preserves the large value of $-\Delta S^{max}$ (143.7 J/kg K for a field change of $\Delta B$ = 5 T at atmospheric pressure for x = 0.40) and, together with the magnetic field, can be used to significantly increase the effective relative cooling power.

Polycrystalline $(MnNiSi)_{1-x}(FeCoGe)_x$ ($x$=0.37, 0.38, 0.39, and 0.40) samples were prepared by arc-melting the constituent elements of purity better than 99.9% in an ultra-high purity argon atmosphere. The samples were annealed under high vacuum for 3 days at 750°C followed by quenching in cold water. Temperature-dependent XRD measurements to determine the crystal structures of the samples were conducted on a Bruker D8 Advance diffractometer using a Cu $K\alpha_1$ radiation source ($\lambda$ = 1.54060 Å) equipped with a LYNXEYE XE detector. Rietveld refinement was used to determine unit cell volumes, and phase fractions above and below the phase transition temperatures using TOPAS. A superconducting quantum interference device magnetometer (SQUID, Quantum Design MPMS) was used to measure the magnetization ($M$) of the $(MnNiSi)_{1-x}(FeCoGe)_x$ samples within the temperature interval of 10-400 K, and in applied magnetic fields ($B$) up to 5 T. Magnetic measurements under applied hydrostatic pressure ($P$) were performed in a commercial BeCu cylindrical pressure cell (Quantum Design). Daphne 7373 oil was used as the pressure transmitting medium. The value of the applied pressure was calibrated by measuring the shift of the superconducting transition temperature of Pb used as a reference manometer (Pb has a critical temperature $T_C$ ~ 7.19 K at ambient pressure) [13]. From the isothermal magnetization [$M(B)$] curves, $-\Delta S$ was estimated using the integrated Maxwell relation, $-\Delta S = \int_0^B \left(\frac{\partial M}{\partial T}\right)_B \mathrm{d}B$. Alternatively, the Clausius-Clapeyron equation, $\frac{\Delta S}{\Delta M} = \frac{\mathrm{d}B}{\mathrm{d}T}$, was employed to calculate the values of $-\Delta S^{max}$ from thermomagnetization curves [$M(T)$] measured at different constant magnetic fields.

The MnNiSi compound undergoes a structural transition from a low-temperature orthorhombic TiNiSi-type structure to a high-temperature hexagonal $Ni_2In$-type structure at an extremely high temperature of about 1200 K in the paramagnetic state, and undergoes a second-order ferromagnetic transition at $T_C$ = 662 K [14, 15]. It is important to tune the transition (and therefore operating temperature of the MCE) so that it occurs near room temperature, a feat that, in this case, could not be accomplished with a single-element substitution. As an alternative substitution strategy, MnNiSi was alloyed with isostructural FeCoGe (having a stable hexagonal $Ni_2In$-type structure and $T_C$ ~ 370 K [16]), which stabilized the hexagonal $Ni_2In$-type phase by sharply reducing the structural transition temperature from 1200 K in MnNiSi to less than 300 K. As a result, coupled MSTs have been realized in this system over a wide temperature range that spans room temperature. The MST in the $(MnNiSi)_{1-x}(FeCoGe)_x$ compounds remains coupled only for $0.40 \leq x \leq 0.37$, but spans a large temperature range of 235 to 355 K as shown in Fig. 1(a).

The application of hydrostatic pressure ($P$) has an effect that resembles that of increasing the concentration ($x$) of FeCoGe, shifting the magnetostructural transition temperature ($T_M$) to lower temperature by about –10 K per kbar of applied pressure ($dT_M/dP$ ~ –10 K/kbar). Reducing the lattice parameter $a_{ortho}$ in the orthorhombic crystal structure (Fig. 1(b)) distorts the geometry of MnNiSi, resulting in a stabilization of the hexagonal phase [17]. Therefore, the shift in $T_M$ with application of pressure is likely associated with a pressure-induced distortion of the orthorhombic lattice that increases the stability of the hexagonal phase. From the pressure-induced shift in $T_M$, and the volume change through the MST as determined from temperature-dependent X-ray diffraction (XRD), we estimated the equivalent average compressibility per unit substitution of FeCoGe to be approximately $7.93 \times 10^{-11}$ $Pa^{-1}$.

As estimated from magnetization isotherms (Fig. 2(a)) using a Maxwell relation, we have observed a large, field-induced isothermal entropy change ($-\Delta S$) in the vicinity of the MST (Fig. 2(b)). Specifically, the $x$ = 0.40 compound has a $-\Delta S^{max}$ = 143.7 J/kg K for a field change of $\Delta B$ = 5 T, which is about 63% of theoretical limit $-\Delta S_{\mathrm{th}}^{\mathrm{max}} = \mathrm{nRln}(2\mathrm{J}+1)$ =

228.4 J/kg K, where J is the total angular momentum of the magnetic ions, R is the universal gas constant, and n is the number of magnetic atoms per formula unit. The observed value of $-\Delta S^{max}$ is the largest reported to date. The experimental results are summarized in Table I and also compared with other well-known giant magnetocaloric materials.

With the application of hydrostatic pressure, the peaks in the $-\Delta S(T)$ curves shift to lower temperatures at a rate (sensitivity) of about $dT_M/dP \sim -10$ K/kbar, but the MCE remains robust over the temperature ranges shown (Fig. 2(b)). There have been pressure-dependent studies on other systems, but most suffer from things such as difficult sample preparation and reproducibility, and large magnetic hysteresis losses [25-27].

The structural entropy change ($-\Delta S_{st}$) associated with the volume change $\Delta V$ has been estimated (for $x$ = 0.40) by employing the Clausius-Clapeyron equation, $\Delta S_{\mathrm{st}} = -\Delta V \left(\frac{\mathrm{d}T_{\mathrm{M}}}{\mathrm{d}P}\right)^{-1}$. The relative volume change $\frac{\Delta V}{V}$ ~ 2.85% was determined from temperature-dependent XRD measurements made just above and below the MST (Fig. 1(b)). The corresponding structural entropy change is $-\Delta S_{st} = 38.7$ J/kg K.

The Clausius-Clapeyron equation is considered to be more reliable than the Maxwell relation for calculating the entropy change near a first-order transition. Applying the Clausius-Clapeyron equation following Ref. 28, for $x = 0.39$ (Fig. 2(c)) we obtain $-\Delta S^{max} \sim$ 70.7 J/kg K ($\Delta B$ = 5 T). This value exceeds those reported (using the Clausius-Clapeyron equation) for all other well-known MCE materials.

The relative cooling power (RCP = $|-\Delta S^{max} \times \delta T_{FWHM}|$, where $\delta T_{FWHM}$ is the full-width at half-maximum of the $-\Delta S$ vs. $T$ plot) of $(MnNiSi)_{1-x}(FeCoGe)_x$ at ambient pressure varies only moderately with composition (Fig. 3(c)), and the material suffers very low magnetic hysteresis losses (Fig. 2(a)). Although $(MnNiSi)_{1-x}(FeCoGe)_x$ exhibits a very large entropy change, more than an order of magnitude larger than that of Gd metal [18], the narrow width of its $-\Delta S(T)$ curve compromises its applicability for magnetic cooling. In principle, the effective range of the working temperature could be extended by introducing a compositional variation in the material (i.e., gradient materials or composites). However, a more sophisticated strategy would be to take advantage of the sensitivity of the transition temperature to applied hydrostatic pressure (~ 10 K/kbar).

Since the large MCE is maintained as the MST shifts in temperature, a radical improvement of the "effective RCP" of the material could be utilized. Theoretically, it has been suggested that the "effective RCP" of a material undergoing a first-order magnetic phase transition can be improved by applying hydrostatic pressure while simultaneously varying the applied magnetic field [29]. In essence, this means that the effective width of $-\Delta S(T)$ increases by an amount equal to the temperature shift with pressure. It should be noted that, by definition, this is not the barocaloric effect [11, 12]. In the case of $(MnNiSi)_{1-x}(FeCoGe)_x$ with $x = 0.40$, applying 1 kbar of pressure along with a field change of $\Delta B = 5$ T, increases the effective RCP by a factor of five. In addition, the working temperature range increases to $\delta T_{FWHM} = 10$ K. Figure 3(b) shows the enhancement of the effective RCP by up to factor of fifteen of the compound with $x = 0.39$ under applied pressures up to 3.69 kbar together with the magnetic field 5 T. Interestingly, the effective temperature range spans room temperature through the freezing point of water, which may be ideal for certain cooling applications. Although there are currently no known magnetic refrigeration prototypes that can utilize this effect, it is something that could be employed in next-generation devices.

In summary, we have shown that combining two isostructural compounds with extremely different magnetic and thermo-structural properties can result in a new system that possesses extraordinary magnetocaloric properties with an acute sensitivity to applied pressure. The magnetic compound, $(MnNiSi)_{1-x}(FeCoGe)_x$, represents a new class of room-temperature magnetocaloric materials that exhibits extraordinarily large magnetocaloric effects and fit many of the criteria for an ideal magnetocaloric material including: (i) it suffers no magnetic hysteresis losses; (ii) it is composed of nontoxic, abundant materials; and (iii) it has a straightforward and repeatable synthesis processes. The characteristic that makes these new materials promising, however, is their response to applied hydrostatic pressure, which provides a means to optimize the magnetocaloric effect at any temperature within its active range.

**Acknowledgements**

Work at Louisiana State University (S. Stadler) was supported by the U.S. Department of Energy (DOE), Office of Science, Basic Energy Sciences (BES) under Award No. DE-FG02-13ER46946, and heat capacity measurements were carried out at LSU by P. W. Adams who is supported by DOE, Office of Science, BES under Award No. DE-FG02-07ER46420. Work at Southern Illinois University was supported by DOE, Office of Science, BES under Award No. DE-FG02-06ER46291. D. P. Young fabricated samples and acknowledges support from the NSF through DMR Grant No. 1306392. XRD measurements were carried out by J. Y. Chan who was supported by NSF under DMR Grant No. 1360863.

**TABLE I.** Transition temperatures ($T_C$ or $T_M$), and observed -$\Delta S^{max}$ for materials exhibiting giant MCE including $(MnNiSi)_{1-x}(FeCoGe)_x$ (present work) for a field variation of 5 T near room temperature.

| Material | $T_C$ or $T_M$ (K) | -$\Delta S^{max}$ (J/kg K) | References |
|---|---|---|---|
| Gd | 294 | 10.2 | [18] |
| $(MnNiSi)_{1-x}(FeCoGe)_x$ | | | [Present work] |
| $x$ = 0.40 | 276 | 143.7 | |
| $x$ = 0.39 | 305 | 85.2 | |
| $x$ = 0.38 | 318 | 87.5 | |
| $(NiMnSi)_{0.56}(FeNiGe)_{0.44}$ | 292 | 11.5 for $\Delta B$ = 1 T | [4] |
| $Mn_{1-x}Cu_xCoGe$ | | | [3] |
| $x$ = 0.08 | 321 | 53.3 | |
| $x$ = 0.085 | 304 | 52.5 | |
| $x$ = 0.09 | 289 | 41.2 | |
| $x$ = 0.095 | 275 | 34.8 | |
| $x$ = 0.1 | 249 | 36.4 | |
| $MnCoGeB_x$ | | | [2] |
| $x$ = 0.01 | 304 | 14.6 | |
| $x$ = 0.02 | 287 | 47.3 | |
| $x$ = 0.03 | 275 | 37.7 | |
| $Mn_{1-x}Cr_xCoGe$ | | | [7] |
| $x$ = 0.04 | 322 | 28.5 | |
| $x$ = 0.11 | 292 | 27.7 | |
| $x$ = 0.18 | 274 | 15.6 | |
| $Mn_{1-x}V_xCoGe$ | | | [8] |
| $x$ = 0.01 | 322 | 8.7 for $\Delta B$ = 1.2 T | |
| $x$ = 0.02 | 298 | 9.5 | |
| $x$ = 0.03 | 270 | 3.4 | |
| $MnCo_{0.95}Ge_{1.14}$ | 331 | 6.4 for $\Delta B$ = 1 T | [9] |
| $Gd_5Si_2Ge_2$ | 272 | 36.4 | [19] |
| MnAs | 318 | 30 | [20] |
| $MnFeP_{0.45}As_{0.55}$ | 305 | 18 | [21] |
| $La(Fe_{0.88}Si_{0.12})_{13}H_1$ | 274 | 23 | [22] |
| $Ni_{55.2}Mn_{18.6}Ga_{26.2}$ | 320 | 20.4 | [23] |
| $Ni_2Mn_{1-x}Cu_xGa$ | | | [24] |
| $x$ = 0.25 | 318 | 64 | |
| $x$ = 0.26 | 309 | 42 | |

**Figure Captions:**

FIG. 1(a) Temperature dependence of the magnetization in the presence of a 0.1 T magnetic field during heating and cooling (direction indicated by arrows) for $(MnNiSi)_{1-x}(FeCoGe)_x$ as measured at ambient pressure (solid lines) and at different applied hydrostatic pressures (broken lines). (b) XRD patterns for $x$ = 0.40 measured at temperatures immediately before and after the magnetostructural transition. The Miller indices of the high-temperature hexagonal and low-temperature orthorhombic phases are designated with and without an asterisk (*), respectively.

FIG. 2(a) The isothermal magnetization curves for $x$ = 0.40. Note the negligible magnetic hysteresis loss (i.e., the magnetization curves are reversible in field) in the vicinity of magnetostructural transition. (b) Plots of the isothermal entropy change ($-\Delta S$) as a function of temperature were estimated using a Maxwell relation for magnetic field changes of $\Delta B$ = 5 T (upper curves) and 2 T (lower curves), measured at ambient pressure (solid lines) and at different applied hydrostatic pressures (broken lines). The "star" symbols inside each $-\Delta S(T)$ curve represents the corresponding total entropy change estimated employing the Clausius-Clapeyron equation for $\Delta B$ = 5 T. A linear fit of these values, intended as a guide to the eye, is indicated by a black dotted line. (c) Heating thermomagnetization curves for applied fields $B$ = 0.1 and 5 T used to estimate the value of $-\Delta S$ for $x$ = 0.39 using the Clausius-Clapeyron equation.

FIG. 3(a) RCP as a function of temperature at ambient pressure for $(MnNiSi)_{1-x}(FeCoGe)_x$ (present work) and other well-known magnetic refrigerant materials: Gd (Ref. 18), $Gd_5Si_2Ge_2$ (Ref. 19), MnAs (Ref. 20), $MnFeP_{0.45}As_{0.55}$ (Ref. 21), and $Ni_2Mn_{0.75}Cu_{0.25}Ga$ (Ref. 24). $T_a$ is the temperature corresponding to $-\Delta S^{max}$ for a field change of 5 T. To highlight the remarkable enhancement in the effective RCP with the application of 1 kbar pressure, a linear fit of the composition-dependent values of the RCP have been made and are indicated by broken lines. (b) The pressure-induced enhancement of the effective RCP has been estimated and is shown for $x$ = 0.39. From the linear fitting of $-\Delta S^{max}$ at ambient pressure and at different applied pressures, we determined the value of $-\Delta S^{max}$ at the midpoint between the $-\Delta S(T)$ peaks at ambient pressure and the highest applied pressure. The full widths at half maximum of $-\Delta S(T)$ ($\delta T_{FWHM}$) at ambient pressure and with application of the largest pressure have been denoted by horizontal lines with the small and large rectangles, respectively. The effective RCP is the area inside the rectangle.

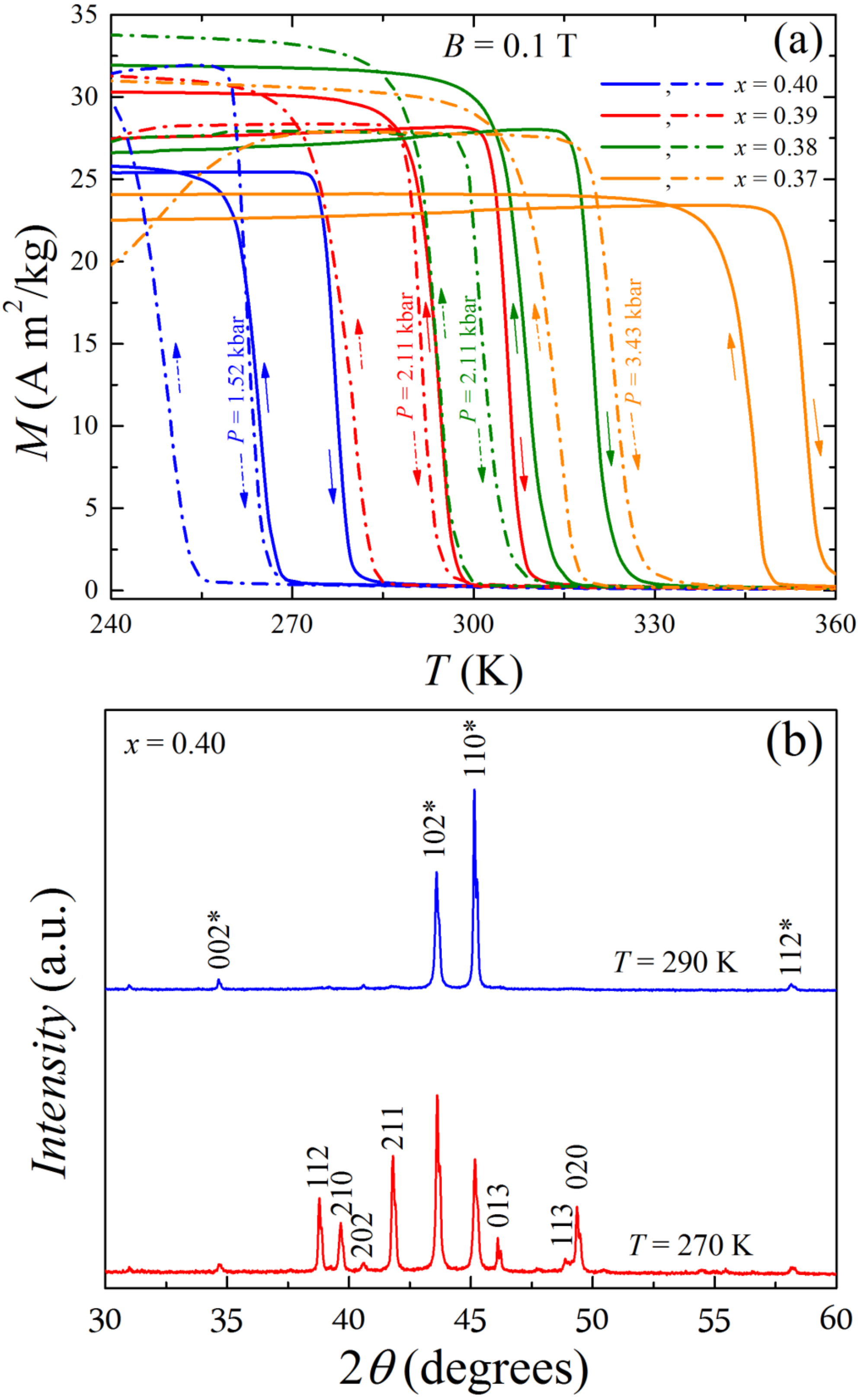

FIG. 1

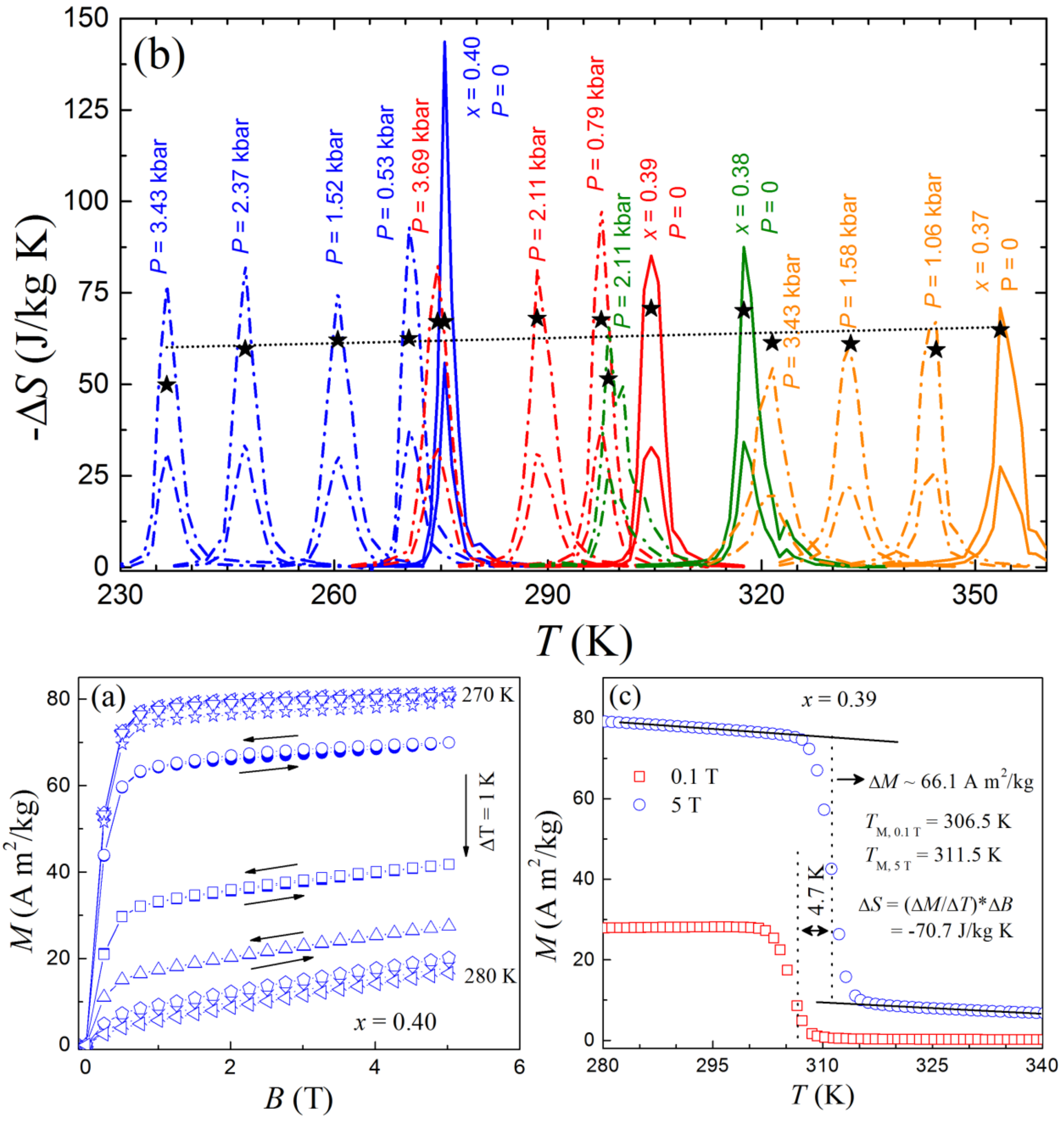
(b)
-ΔS (J/kg K)
T (K)
P = 3.43 kbar
P = 2.37 kbar
P = 1.52 kbar
P = 0.53 kbar
P = 3.69 kbar
x = 0.40
P = 0
P = 2.11 kbar
P = 0.79 kbar
P = 2.11 kbar
x = 0.39
P = 0
x = 0.38
P = 0
P = 3.43 kbar
P = 1.58 kbar
P = 1.06 kbar
x = 0.37
P = 0
(a)
270 K
ΔT = 1 K
280 K
x = 0.40
M (A m²/kg)
B (T)
(c)
x = 0.39
0.1 T
5 T
ΔM ~ 66.1 A m²/kg
T_M, 0.1 T = 306.5 K
T_M, 5 T = 311.5 K
4.7 K
ΔS = (ΔM/ΔT)*ΔB
= -70.7 J/kg K
T (K)

FIG. 2

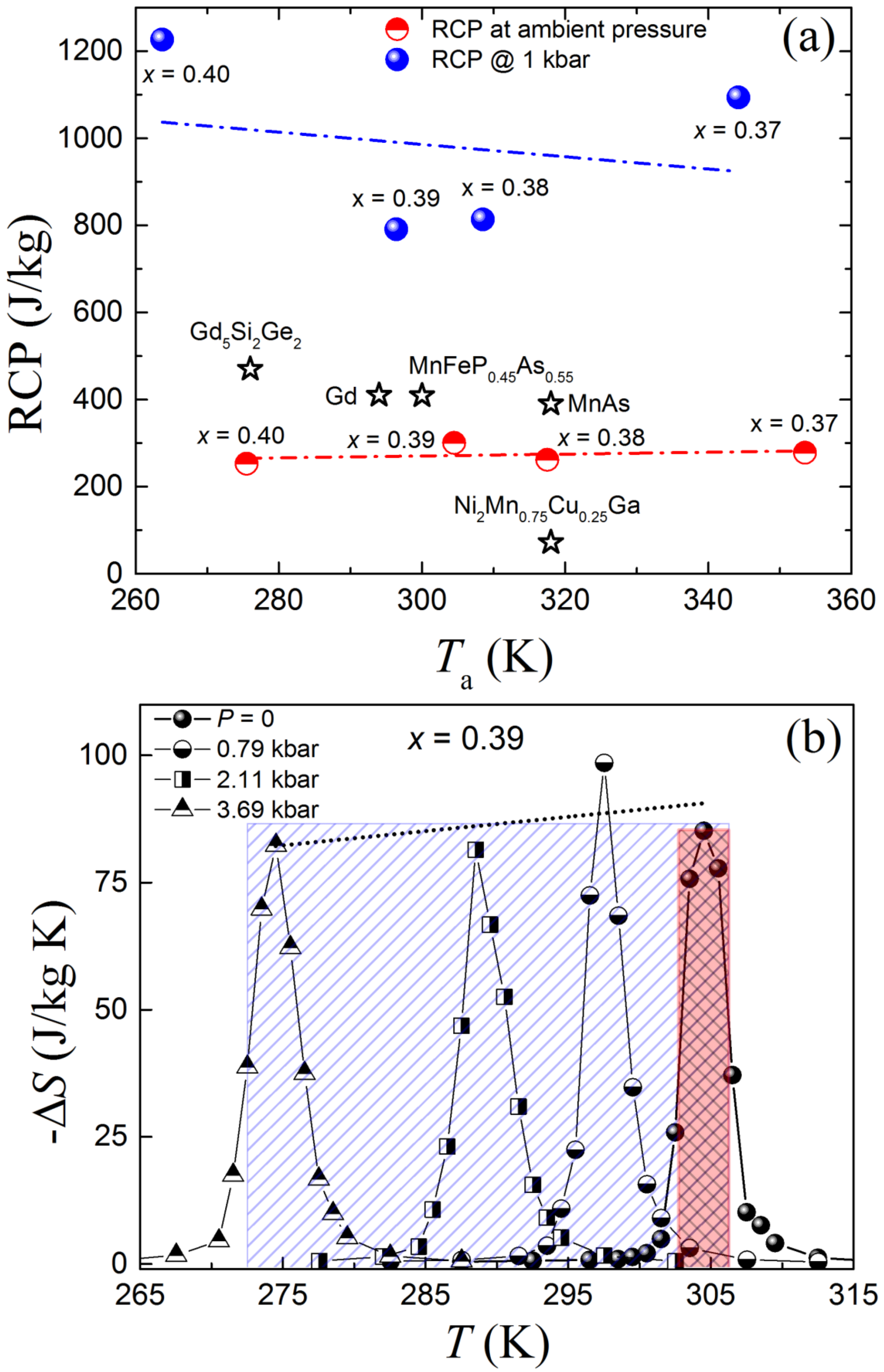

FIG. 3